# REVIEW AND ANALYSIS OF THERMOPHYSICAL PROPERTIES OF SULFURIC ACID–WATER ELECTROLYTE

L. Oca[a £], Jose Miguel Campillo-Robles[b #], M. Mounir Bou-Ali[c *]

[a]*Electronics and Computing Department, Faculty of Engineering, Mondragon Unibertsitatea, 20500 Arrasate, Basque Country, Spain.*

[b]*Elektrokimikako ikerketa-taldea, Mekanika eta Ekoizpen Industrialeko Saila, Mondragon Unibertsitatea, Loramendi 4, 20500 Arrasate, Basque Country.*

[c]*Mechanical and Industrial Product area, Faculty of Engineering, Mondragon Unibertsitate, 20500 Arrasate, Basque Country, Spain.*

**Corresponding author (M. Mounir Bou-Ali): Tel.: +(34) / +(34) 627862858*

*E-mail addresses: [£]lauraoca@mondragon.edu; [#]jmcampillo@mondragon.edu; [*]mbouali@mondragon.edu*

## Abstract

In this work, we have performed a critical review of thermophysical properties of sulfuric acid water mixtures. The thermophysical properties analysed are: density, viscosity, refraction index, thermal expansion coefficient and mass expansion coefficient. Density of sulfuric acid-water mixture has been measured for different mass fractions (0.1 to 0.4 *w*) in a broad temperature range (273.15 to 333.15 K). A new parameterization of the density has been performed using a least-square method. In this parameterization, the number of coefficients has been reduced with respect to previous works. Moreover, this new equation has been analysed in an extended region of temperatures and concentrations in ranges: {*w* (0 to 0.1); *T* (233.15 to 373.15 K)}. The coefficients of thermal and mass expansion have also been calculated experimentally and a linear relation of the density has been determined in function of the variation of the temperature and concentration. In addition, viscosity and refraction index measurements have been performed at 298.15 K and 293.15 K respectively. Our results have been checked measurements of the literature (review), and show good correspondence.

## 1. Introduction

Sulfuric acid has been known and used since the Middle Ages. It has many desirable properties that lead to its use in a wide variety of applications.[1] For this reason, it has been an important item of commerce since the 17th century.[2] Nowadays, sulfuric acid is one of the most important inorganic compounds produced by the chemical industry.[3-5] In fact, sulfuric acid is used as a raw material or as a processing agent in the production of nearly all manufactured goods. The major use of sulfuric acid is in the production of fertilizers and in the manufacture of chemicals, but it is also needed for batteries, explosives, derived products of petroleum, detergents, dyes, insecticides, drugs, plastics, steel, and many other products.[6] Due to the sulfuric acid demand of emerging countries, the worldwide production is in continuous increase.[7] The world production of sulfuric acid is estimated to be greater than 260 million tonnes, and all the forecasts show that it will continue increasing.[8] Moreover, sulfuric acid consumption is often cited as an indicator of the general state of the economy of one country.[9]

The properties of sulfuric acid-water mixtures have been studied systematically from the 19th century.[10-11] Therefore, there is an impressive mass of experimental evidence. However, some physicochemical properties of these mixtures are still being studied.[12-15] At this moment, the properties of these mixtures are been studied deeply due to the use of the sulfuric acid as an electrolyte in lead-acid batteries and its role in atmospheric processes.

Nowadays, the research performed to develop energy storage systems is principally focused on getting advanced batteries,[16-18] such as advanced lead-acid batteries.[19-21] The electrolyte of this kind of battery is composed of sulfuric acid and water, and the performance and useful life of this battery are affected by the properties of this mixture.[22-23] Today, additives for the electrolyte of lead-acid battery are being studied, for the improvement of the thermal physical properties of the electrolyte and the battery useful life.[24-27] Moreover, small amounts of impurities could affect severely the performance and lifetime of this device.[28]

Furthermore, the scientific community notices the importance of sulfuric acid in atmospheric processes. Nowadays atmospheric nucleation processes are being analysed.[29-30] This strong acid plays a key role in the formation of aerosols in the troposphere and the stratosphere.[31-32] Sulfuric acid aerosols influence the quality of life through its climatic and health effects. These aerosol particles participate in the formation of acid rain, which pollutes the soil and the ground water courses of the earth.[33] Sulfuric acid aerosols also modify the radiative balance of the atmosphere by scattering and absorbing sunlight.[34] This process affects the global temperature and thus the global climate. Indeed, at this moment, research is focused on the possibility of conditioning the $CO_2$ greenhouse effect by means of these aerosols.[35]

The main objective of this work is to compare and contrast experimental results of the thermophysical properties of the sulfuric acid-water mixture with measurements of this work, and provide an historic overview of those properties. To reach that goal, experimental density, coefficient of thermal expansion, coefficient of mass expansion, dynamic viscosity and refraction index values have been measured. For checking the previous results, we have performed a systematic measurement of the thermophysical properties previously analysed of these mixtures as a function of temperature and concentration. Moreover, density parametrizations of the literature have been analysed, and we have proposed a new simple parameterization based on our experimental results, that could be useful for modeling of different systems.

## 2. Density

Sulfuric acid-water electrolyte density has been measured extensively, in a wide temperature and concentration range. As a result, there is a great amount of experimental data available. Three review works attempted to order and compare those values of density measurements.[36-38] In 1960, Timmerman collected data from 19 density measurements, and cited another 34 works dismissing them.[36] Later, in 1975, Potter et al. collected 11 measurements in a bibliographic review of components of geothermal waters.[38] Finally, in 2011, Clegg *et. al.* published another review with 16 density measurements.[37] These two review works are useful to make a quick view of the experimental literature. All in all, we believe that a wider comparison between the most remarkable measurements is needed. Therefore, in this document we present a systematic review of the most cited works of the density of sulfuric acid-water electrolyte.

### 2.1. Critical review of density

In order to perform the selection of data available in the bibliography, some criteria have to be defined. On the one hand, we have rejected data before 1900, due to insufficient accuracy or because experimental methods and units are confusing. This criterion agrees with Timmermans[36]. In addition, measurements discarded by Timmermans[36] in 1960, as the author claims that these data have only historical value and they are not accurate enough. On the other hand, measured density values have been included. Added to that, authors that have taken values from other sources have been identified. Moreover, reviews of measurements that other authors consider of great importance have also been taken into account. Table 1 collects the most remarkable information about the measurements of the bibliography.

**Table 1.** Density measurements of the sulfuric acid-water solutions in the literature chronologically ordered, where $T$ is Temperature, $w$ is the Mass Fraction and $\rho$ is the density. The author's highlighted in grey corresponds to the data used for the comparison of our experimental work and fitting. NI: No information found.

| First author | Year | T/K [number of values or $\Delta T$ from previous temperature in K] | $w$ [number of values or $\Delta w$ in %] | Measured magnitude *1 | Unit | Measured/ Source *2 |
|---|---|---|---|---|---|---|
| Ferguson[39] | 1904 | 288.75 | 0 to 0.9319 [29] | $\rho(T, 288.75\text{K})$ | - | Yes |
| Domke[40] | 1905 | 273.15 to 333.15 [10 K] | 0 to 1 [1 %] | $\rho(T, 288.15\text{K})$ | - | Yes |
| Börnstein[41] | 1912 | 288.15, 298.15 | 0 to 1 [1 %] | $\rho(T, 277.15\text{K})$ | - | No / Domke. |
| (A) Sullivan[42] | 1918 | 273.15 to 323.15 [10 K] | 0 to 0.9319 [73] | $\rho(T, 288.75\text{K})$ | - | No / Ferguson |
| (B) Sullivan[42] | 1918 | 278.15, 288.15, 298.15 | 0 to 0.9319 [29] | $\rho(T, 288.75\text{K})$ | - | No / Ferguson |
| Rhodes[43] | 1923 | 288.75 | 0 to 0.9960 [20] | $\rho(T)$ | g·cm$^{-3}$ | Yes |
| Washburn (editor)[44] | 1928 | 288.75 | 1 to 1 [1 %] | $\rho(T, 277.15\text{K})$ | - | No / Domke and different sources. |
| Forsythe (editor)[45] | 1954 | 273.15 to 348.15 [25 K]<br>273.15 to 333.15 [10 K]<br>288.15, 298.15, 353.15, 373.15 | 0 to 0.98 [1 %]<br>0.4, 0.7<br>0.3 to 0.7 [10 %] | $\rho(T, 277.15\text{K})$ | - | No / Domke |
| (A) Fasullo[46] | 1965 | 293.15<br>233.15<br>244.25, 255.35 | 0.2 to 0.7 [10 %]<br>0 to 0.9 [10 %]<br>0 to 1 [10 %] | $\rho(T)$ | lb·ft$^{-3}$ | NI |
| (B) Fasullo[46] | 1965 | 266.45 | 0 to 0.6498 [53]<br>0.65 to 1 [0.1 %] | $\rho(T, 288.75\text{K})$ | - | No / Ferguson |
| Haase[47] | 1966 | 277.55<br>288.65 to 422.05 | 0.03929 to 0.7706 [16]<br>0.03929 to 0.8593 [19] | $\rho(T)$ | kg·L$^{-1}$<br>- | No / Washburn |
| (A) D´Ans[48] | 1967 | 288.75 | 0.02 to 1 [5 %] | $\rho(T)$ | g·cm$^{-3}$ | No / Washburn |
| (B) D´Ans[48] | 1967 | 273.15 | 0.05 to 1 [12] | $\rho(T)$ | g·cm$^{-3}$ | No / Rhodes |
| Bode[49] | 1977 | 273.15 to 323.15 [10 K], 298.15 | 0 to 0.5 [ 2 %]<br>0.5 to 1 [5 %] | $\rho(T, 277.15\text{K})$ | - | No / Washburn |
| (A)Weast (editor)[50] | 1981 | 273.15 to 303.15 [10 K] | 0 to 1 [1 %] | $\rho(T, 277.15\text{K})$ | - | No / Washburn |
| (B)Weast (editor) [50] | 1981 | 288.15, 298.15, 323.15, 348.15, 373.15 | 0 to 0.1 [0.5 %]<br>0.1 to 0.2 [1 %]<br>0.2 to 1 [2 %] | $\rho(T, 277.15\text{K})$<br>$\rho(T, 293.15\text{K})$ | - | No / Washburn |
| Kaye[51] | 1986 | 273.15 to 348.15 [25 K] | 0.01731 to 1 [65] | $\rho(T)$ | kg·m$^{-3}$ | NI |
| Maksimova[52] | 1987 | 273.15 to 323.15 [25 K] | 0.1 to 0.3 [10 %]<br>0.3 to 0.8 [5 %] | $\rho(T)$ | g·cm$^{-3}$ | Yes |
| Beyer[53] | 1996 | 293.15<br>293.15 | 0.389, 0.630<br>0.523, 0.886 | $\rho(T)$ | -<br>g·mL$^{-1}$ | Yes |
| (A) Aseyev[54] | 1996 | 293.15 | 0.3 to 0.76 [2 %] | $\rho(T)$ | kg·m$^{-3}$ | No / Different sources |
| (B) Aseyev[54] | 1996 | 233.15 to 283.15 [10 K]<br>323.15, 343.15, 363.15 | 0.02 to 1 [2 %]<br>0.04 to 1 [2 %] | $\rho(T)$ | kg·m$^{-3}$ | NI |
| (C) Aseyev[54] | 1996 | 195.15 | 0.1 to 0.8 [2 %] | $\rho(T)$ | kg·m$^{-3}$ | NI |
| Perry[55] | 2008 | 221.15 | 0.01 to 1 [1 %] | $\rho(T, 277.15\text{K})$ | - | No / Washburn |
| Myhre[56] | 2003 | 233.15 to 273.15 [5 K]<br>273.15 to 318.15 [5 K]<br>323.15 to 373.15 [5 K]<br>273.15 to 363.15 [5 K]<br>273.15 to 333.15 [10 K]<br>288.15, 298.15, 353.15, 373.15 | 0.123<br>0.291<br>0.503<br>0.585<br>0.672<br>0.765 | $\rho(T)$ | kg·m$^{-3}$ | Yes |
| Salkind[57] | 2002 | 262.15 to 300.15 [13] | 0 to 0.597 [11] | $\rho(T, 288.15\text{K})$ | - | No / Domke |
| Pavlov[58] | 2017 | 241.15 to 303.15 [14] | 0 to 0.5 [2 %]<br>0.5 to 0.6 [5 %] | $\rho(T, 277.15\text{K})$ | - | No / Washburn |
| Liu[59] | 2012 | 221.15 to 301.15 [13] | 0.0746 to 0.3994 [22] | $\rho(T, 298.15\text{K})$ | - | Yes |

*1 Density $\rho(T)$ and relative density $\rho(T, 288.75\text{K})$ are differentiated, as relative density represents a unitless value in relation to a density at a certain temperature (i.e. 288.15 K).

[*2] Clarification of which works have performed original measurements and which ones only referred other measurements are presented. Where *yes* means direct measurements and *no/source* explain from witch source are taken the presented values.

In order to compare the literature values with our work, authors highlighted in grey in Table 1 have been selected. Five references have been roughly compared with our experimental values. First of all, Washburn[44], as many authors take this book as a reference.[47-50] In this work, values are mostly taken from Domke and Bein[40] although values at 353.15 K and 373.15 K have been taken from different sources. Secondly, Fasullo[46], Maksimova et. al.[52] and Myhre *et al*.[56] measurements have been analysed, as they have values at low and high temperatures (below 273.15 K and over 333.15 K and with a wide range of concentrations). Last but not least, Liu and Li measurements have been compared, as they have characterized lead-acid battery electrolyte.[59]

### 2.2. Density parametrization

The parameterization of the density data of liquids as a single function of the concentration and the temperature is a key factor from a practical point of view. This kind of equations is often needed in numerical algorithms of chemical engineering and physical chemistry. Therefore, different kind of parametrizations of the density have been developed in the literature.[60] One example of technological importance is the electrolyte of the lead-acid batteries. In this practical case, the temperature and density of the electrolyte are the measurable parameters. A numerical relation between concentration and these two parameters is necessary, because some of the governing equations of the battery electrochemistry are functions of the electrolyte concentration.[61-64]

In this section, first, a bibliography research on parametrization of sulfuric acid-water mixture has been done. Then, a comparison of our experimental data and with the literature has been performed. Finally, a new and simple parametrization based on our experimental data has been proposed.

#### 2.2.1. Review of parametrizations

The identified parametrizations are collected in Table 2.

**Table 2.** Parametrizations of the density of sulfuric acid-water solutions in the literature chronologically ordered (A, B, C are coefficients used to adjust the equation). Where *T* is Temperature, *w* is the Mass Fraction and *ρ* is the density, *x* corresponds to the molar fraction (with units of mol dm$^{-3}$ in Novotný[65] and mole % in Walrafen[68].

| First author | Year | *T*/K | *w* | Expression | *ρ* | *T* | Number of coefficients |
|---|---|---|---|---|---|---|---|
| Novotný[65] | 1988 | 273.15 to 373.15 | 0 to 1 | $\rho = \rho_{water}(T) + Ax + BxT + CxT^2 + Dx^{3/2} + Ex^{3/2}T + Fx^{4/2}T^2$ | kg m$^{-3}$ | °C | 21 |
| Aseyev[54] | 1996 | 273.15 to 363.15 | 0.1 to 0.8 | $\rho = \rho_{water}(T) + Aw + BwT + Cw^2$ | kg m$^{-3}$ | °C | 11 |
| Myhre[66] | 1998 | 273.15 to 323.15 | 0.1 to 0.9 | $\rho = \sum_{i=0}^{10}\sum_{j=0}^{4} A_{ij} w^i (T - 273.15)^j$ | kg m$^{-3}$ | K | 35 |
| Kulmala[67] | 1998 | 233.15 to 298.15 | 0.1 to 1 | $\rho = \rho_1 + (\rho_2 - \rho_1)\frac{(T - 273.15)}{60}$ <br> $\rho_1 = A + Bw - Cw^2 + Dw^3 - Ew^4; \rho_2 = F + Gw - Hw^2 + Iw^3 - Jw^4$ | kg m$^{-3}$ | ºC | 8 - 10 |
| Walrafen[68] | 2000 | 293.15 | 0.7 to 1 | $\rho = A + Bx + Cx^2 + Dx^3 + Ex^4 + Fx^5$ | g cm$^{-3}$ | ºC | 5 |
| Vehkamäki[69] | 2002 | 273.15 to 373.15 | 0 to 1 | $\rho = A(w) + B(w)T + C(w)T^2$ | g cm$^{-3}$ | K | 21 |
| Myhre[56] | 2003 | 273.15 to 323.15 | 0.1 to 0.9 | $\rho = \sum_{i=0}^{10}\sum_{j=0}^{3} A_{ij} w^i (T - 273.15)^j$ | kg m$^{-3}$ | K | 32 |
| Hyvärinen[70] | 2005 | 297.35 | 0 to 1 | $\rho = \rho_{water} + Aw + Bw^2 + Cw^3 + Dw^4$ | kg m$^{-3}$ | ºC | 4 |
| Clegg[37] | 2011 | 273.15 to 373.15 | 0.06 to 1 | $\rho(T,w) = \rho(T_r,w) + (T - T_r)(Q_1(T_r,w)) - T_r Q_2(T_r,w) + Q_2(T_r,w)(T^2 - T_r^2)/2$ | g cm$^{-3}$ | K | 95 |

Novotný and Söhnel justified an empirical equation for describing the density of binary aqueous solutions.[65] This equation is a polynomial expansion with 6 adjustable constants and one term related to the density of the water. This term is an empirical equation of temperature that has three non-adjustable parameters. Novotný and Söhnel tested the validity of the equation for more than 300 binary aqueous solutions. In the case of sulfuric acid-water mixtures, they fitted data of Domke and Bein[40], Washburn[44], and D´Ans and Lax[48]. Due to the special behaviour of the mixture, they performed three different fits between 273.15 K and 373.15 K in these concentration ranges: (0 to 0.7 *w*), (0.71 to 0.9 *w*) and (0.91 to 1 *w*). Later, Aseyev and Zaytsev[54] presented another polynomial expansion with three adjustable constants, and one term related to the density of the water. This last term is a function of temperature that has nine non-adjustable parameters and is valid in the region {*w* (0.1 to 0.8); *T* (273.15 to 363.15 K)}. Author's take the experimental data from Domke and Bein[40], Washburn[44], and D´Ans and Lax[48]. In 1998, Myhre *et al*. performed another polynomial fitting of the density of the sulfuric acid-water mixtures for the {*w* (0.1 to 0.9); *T* (210.15 to 323.15 K)} region.[66] They fitted their measurements of density at low temperatures (see Table 2) and the data of Washburn[44]. The parameterization developed by Myhre *et al*. had 35 adjustable terms. However, this equation contained errors resulting in minor inaccuracies at low temperatures. For this reason, afterwards, in 2003, Myhre and coworkers improved their fitting of the data. They obtained an equation with 32 adjustable constants for the same concentration and temperature ranges[56]. Kulmala *et.al.*[68] provide another parametrization taken from Jaecker-Voirol[71].

Walrafen[68] fitted with a fifth-degree least-square polynomial the density values for the range of 70 – 100 mole% of Kaye and Laby[51] with 5 constants. Vehkamäki *et al*. presented another parameterization of the density[70]. In this case, they used only 21 adjustable constants for fitting the densities of Washburn[44]. This parameterization reproduced well the low temperature (220.15 to 300.15 K) densities of Myhre *et al*.[56]. Hyvärinen *et. al.*[71] studied densities of ternary $H_2SO_4$ + $NH_3$ + $H_2O$ solutions. Sulfuric acid water mixture regression can be determined when $NH_3$ is zero at constant temperature of 297.35 K. Last but not least, Clegg *et. al.*[37] provided a fitting equation with 95 terms to adjust. All in all, this fitting can accurately respond in the whole concentration and temperature range.

**2.2.2. Density measurements**

The sulfuric acid-water mixtures analysed in the present work were prepared using the chemical sample given in the Table 3.

**Table 3.** Chemical sample information.

| Chemical Name | Source | Initial Mole Fraction Purity | Purification Method | Final Mole Fraction Purity | Analysis Method |
|---|---|---|---|---|---|
| Sulfuric acid | Merck | 0.95 to 0.97 | | ? | - |
| Distilled water | Water still Pobel 909 | ? | Bi - distillation | - | - |

All mixtures are prepared using Gram VXI 310 precision balance of ± 0.01 mg. Although in most of the mixtures the densest liquid is firstly added, in this case, sulphuric acid is added to water, in order to avoid splatters. The dissolution reaction of sulfuric acid in water is highly exothermic, so it has been necessary to cool it during the mixing process, and increase the preparation and relaxation time. We have prepared mixtures with mass fraction ranging from 0.1 to 0.4, mass fraction of densest component. In this range, we have prepared 17 samples, with a difference of concentration between them of approximately 0.02 *w*. Before the measurements, all the samples have been shaken vigorously, and then an ultrasonic bath with temperature control has been applied for degassing.

The density of the sulfuric acid-water mixtures has been measured using a vibrating quartz U-tube density meter, Anton Paar DMA 5000. The density determination is based on measuring the period of oscillation of the vibrating U-shaped tube which is filled with the sample liquid.[72] This density meter measures the density with an accuracy of $\pm 5 \times 10^{-3}$ kg $m^{-3}$, and the temperature with and accuracy of ± 0.01 K. The temperature control in this device is carried out accurately by means of Peltier effect. The density meter was calibrated using air and water at 293.15 K following the procedure of the density meter. The density has been measured three times for each concentration and temperature, and then, the average density has been calculated. The temperature of the measurements has ranged from 273.15 K to 333.15 K, with a difference of temperature between each measurement of 283.15 K. In addition, the density at 298.15 K has been measured, because it is a reference temperature in the battery world[72]. We have measured the density of sulfuric acid-water mixtures in the next temperature and concentration range, {*w* (0.1 to 0.4); $T$ (273.15 to 333.15 K)}. The experimental density values of this work are shown in Table 4.

**Table 4.** Measured density values of the sulfuric acid-water solutions $\rho$, at different Temperatures $T$ and mass fractions $w$.

| | $\rho$/kg m$^{-3}$ | | | | | | | |
|---|---|---|---|---|---|---|---|---|
| $w$ | 273.15 K | 283.15 K | 293.15 K | 298.15 K | 303.15 K | 313.15 K | 323.15 K | 333.15 K |
| 0.097 | 1069.26 | 1066.08 | 1062.30 | 1060.23 | 1058.04 | 1053.60 | 1049.23 | 1043.53 |
| 0.115 | 1085.84 | 1082.07 | 1077.83 | 1075.55 | 1073.22 | 1068.89 | 1063.56 | 1058.71 |
| 0.137 | 1101.81 | 1097.50 | 1092.86 | 1090.39 | 1087.86 | 1082.60 | 1078.45 | 1072.65 |
| 0.154 | 1114.20 | 1109.50 | 1104.52 | 1101.94 | 1099.31 | 1093.86 | 1089.88 | 1084.63 |
| 0.173 | 1128.95 | 1123.83 | 1118.46 | 1115.73 | 1112.95 | 1107.33 | 1101.93 | 1095.88 |
| 0.192 | 1145.91 | 1140.32 | 1134.59 | 1131.69 | 1128.88 | 1123.78 | 1118.59 | 1113.22 |
| 0.211 | 1159.42 | 1153.51 | 1147.52 | 1144.52 | 1141.49 | 1135.36 | 1129.37 | 1124.62 |
| 0.230 | 1177.19 | 1170.92 | 1164.60 | 1161.54 | 1158.39 | 1152.03 | 1148.50 | 1140.31 |
| 0.241 | 1185.57 | 1179.14 | 1172.69 | 1169.48 | 1166.27 | 1160.54 | 1154.05 | 1147.47 |
| 0.269 | 1209.44 | 1201.63 | 1194.85 | 1191.58 | 1188.22 | 1181.52 | 1174.82 | 1168.95 |
| 0.288 | 1224.11 | 1217.14 | 1210.19 | 1206.76 | 1203.33 | 1196.60 | 1189.76 | 1183.88 |
| 0.307 | 1241.05 | 1233.47 | 1227.15 | 1223.63 | 1220.11 | 1213.75 | 1206.84 | 1200.67 |
| 0.327 | 1256.77 | 1249.51 | 1242.27 | 1238.85 | 1235.29 | 1228.03 | 1221.02 | 1214.92 |
| 0.346 | 1272.18 | 1264.81 | 1257.51 | 1253.89 | 1250.30 | 1243.38 | 1236.27 | 1230.89 |
| 0.365 | 1288.04 | 1280.59 | 1273.21 | 1269.57 | 1265.95 | 1258.73 | 1251.57 | 1244.45 |
| 0.384 | 1305.68 | 1298.19 | 1290.73 | 1287.05 | 1283.37 | 1276.20 | 1268.99 | 1263.24 |
| 0.400 | 1319.14 | 1311.55 | 1304.07 | 1300.36 | 1296.68 | 1289.37 | 1282.41 | 1275.20 |

The experimental values of the density obtained in this work have been compared with those authors selected from Table 1. The results are in good agreement with the previous results of the other researchers in all the measured concentration and temperature range. For example, Figure 1 shows all these experimental density values at 273.15, 298.15 and 323.15 K.

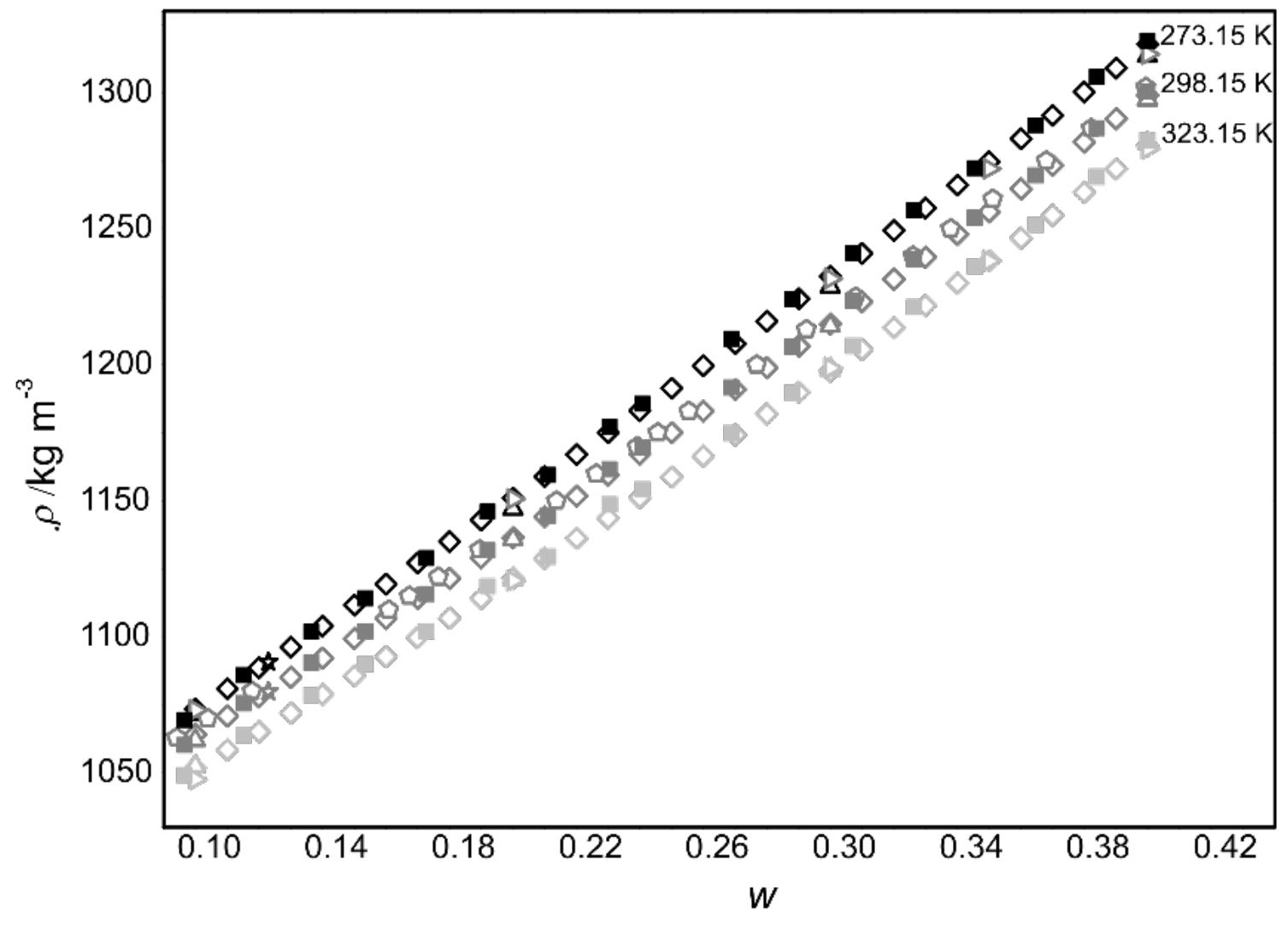

**Figure 1.** Comparison of experimental density values *ρ* obtained in this work in function of mass fractions *w* and bibliography at temperatures *T* of 273.15, 298.15 and 323.15 K. Data from Fasullo[49] is at 277.55, 299.85 and 322.05 K. ◇ Washburn[44], △ Fasullo[46], ▷ Maksimova[52], ⬠ Liu[59], ★ Mhyre[66] and ■ This work.

The experimental values of density of sulfuric acid solution have been found to vary linearly with temperature in the temperature range of measurement. Densities of values from Fasullo[46] have been compared with ours and other authors, however, the temperatures given by Fasullo are 277.55, 299.85 and 322.05 K. In all the cases, the deviation is less than 1 % in the density values of this research and the literature.[44, 46, 52, 59, 66]

### 2.2.3. New density parametrization

The density of sulfuric-acid water electrolyte can be parametrized with a polynomial expansion of the form:

$$\rho\left(w,T\right)=\sum_{i=0}^{n}\sum_{j=0}^{m}\psi_{i,j}w^{i}T^{j} \quad (1),$$

where *T*/K and *w* are the temperature and the concentration/mass fraction of the sulfuric acid in the mixture and $\psi$ are and fitting parameters of the mixture.

The density values measured in this work have been fitted in the region {*w* (0.1 to 0.4); *T* (273.15 to 333.15 K)} using an unweighted least-square method. We have described the experimental densities of the electrolyte using equation (1) with a polynomial expansion of six adjustable parameters of Table 5.

**Table 5.** Coefficients of the polynomial fitting of the density of the sulfuric acid-water solutions at different temperatures and concentrations. Goodness of fit: SSE: 128.64, R-square: 0.9998, RMSE: 0.9948.

| $\psi_{i,j}$ | $j = 0$ | 1 | 2 |
|---|---|---|---|
| $i = 0$ | 1122 (1077, 1168) | - 0.5076 (- 0.8040, - 0.21120) | 2.484 $10^{-4}$ (-2.383 $10^{-4}$, 7.349 $10^{-4}$) |
| 1 | 976.4 (945.4, 1008) | - 1.015 (- 1.111, - 0.9192) | - |
| 2 | 237.8 (215.9, 259.6) | - | - |

We have tested the validity of the fitted equation in our measured range of concentrations and temperatures. This new parameterization describes the experimental results obtained for the density of the electrolyte with a maximum absolute deviation of 2.54 kg m$^{-3}$ and average absolute deviation of 57.6 kg m$^{-3}$. The maximum percent deviation is ±0.225 %. Finally, we have applied the root-mean-square deviation (rmsd) equation obtaining a value of 1 kg m$^{-3}$. All these values confirm the validity of the fitted equation for the accurately description of the density in the measured range of temperatures and concentrations. This can also be seen in Figure 2, where the fitted equation has been plotted (plane) with the measured densities (points).

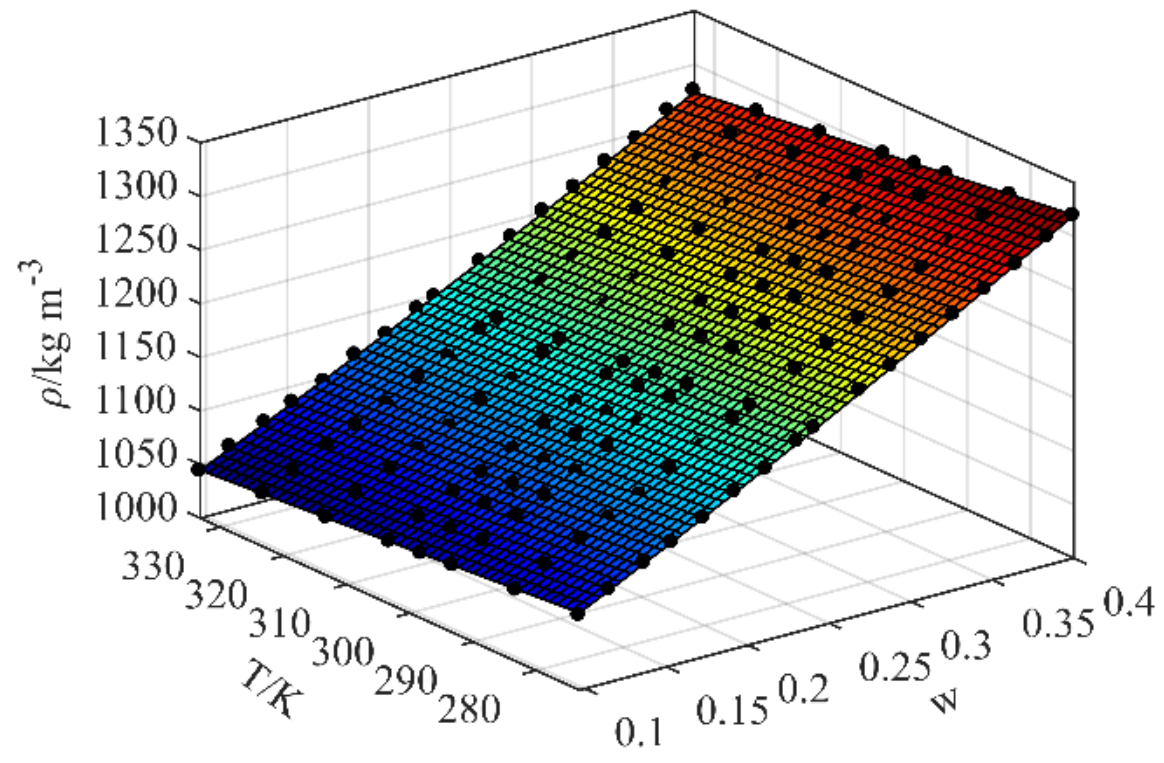

**Figure 2.** Experimental density values of the sulfuric acid-water solutions $\rho$ (points) and least-square fitting of the data (surface) based on eq (1). at different Temperatures $T$ and mass fractions $w$.

The Figure 2 shows a plane in the analysed temperature $T$ and Mass fraction $w$. The density of an specific sample (constant $w$) shows a nearly linear decrease with increasing temperature. At constant temperature, the density of the electrolyte increases with the concentration of acid, showing a nearly linear relation too.

Moreover, our fitting is in concordance with the available fittings in the literature in the measured concentration and temperature range. Further analysis has been done in section 2.3 to set the validity of the equation out of the measured range.

### 2.3. Extrapolation of the proposed fitting equation

In order determine the performance of our parameterization out of the measured range, we have extrapolated the polynomial formula fitted previously (equation (1) and Table 5); and we have compared it with the experimental values of literature (see Table 1). We have analysed the correspondence between the fitted equation and the experimental data in the full temperature (233.15 to 373.15 K) and concentration range (0 to 1 *w*).

#### 2.3.1. Low temperature range, 233.15 < *T* < 273.15 K

In low concentration range (0 to 0.1 *w*), there are no experimental values of the density (see Table 1). That is because, in concentration near to zero (mostly water) the freezing point is near to 273.15 K, so the densities cannot be obtained. At 0.1 *w* concentration, the freezing point is 268.15 K.[73]

In the range of 0.1 to 1 *w*, there are three works with experimental data of the density.[49, 55, 59] Myhre *et al.*[59] measured the density at six specific concentrations of the electrolyte. For example, the experimental densities of Myhre *et al.*[59] and the fitted equation at a concentration of 0.291 *w* have been plotted in Figure 3. For concentrations of 0.123 and 0.291 *w* the values of the density obtained with the fitted equation show a maximum percent deviation with respect to the experimental values of less than 1 % for all the measured temperature range.

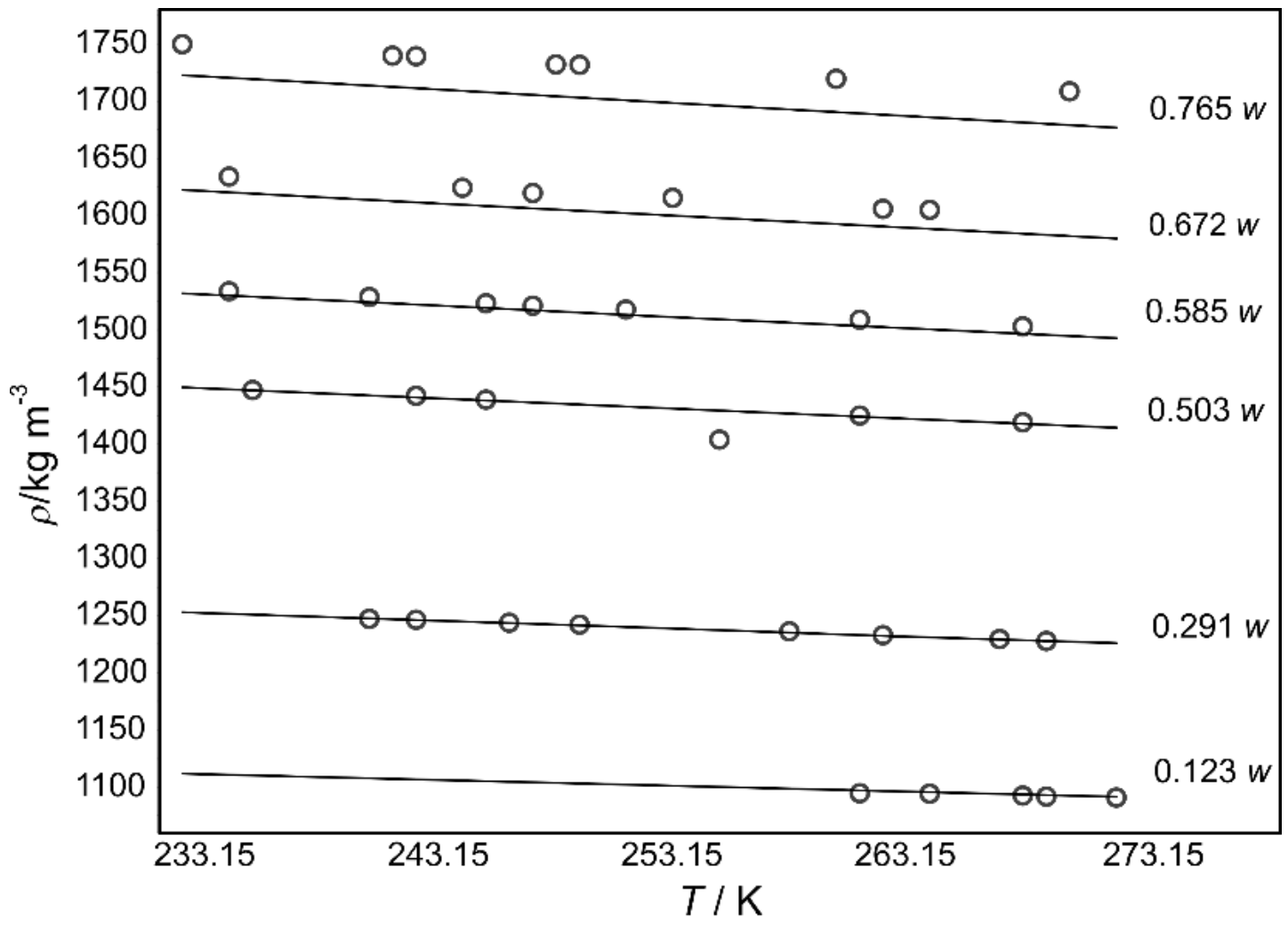

**Figure 3.** Extrapolation of the density parametrization: density values $\rho$ at low temperature range $T$ (233.15 to 273.15 K) for mass fractions $w$ of (0.123 to 0.765). ○ Myhre[66] and - This work, Eq. (1).

Taking into account values measured by Myhre *et al*. we validated our fitted equation {$w$ (0.503 to 0.672); $T$ (221.15 to 273.15 K)} as the maximum percent deviation is less than 1 %. In the highest mass fractions, 0.765 $w$, the maximum percent deviation with respect to the values of Myhre *et al*. is greater, but less than 2 %. Figure 3 shows the fitted equation and the experimental values at 273.15 K.

### 2.3.2. Medium temperature range, 273.15 < $T$ < 323.15 K

At low mass fractions (0 to 0.1 $w$) the fitting equations is in good agreements with results of the bibliography,[44, 46, 52, 59] as it is shown in Figure 4.

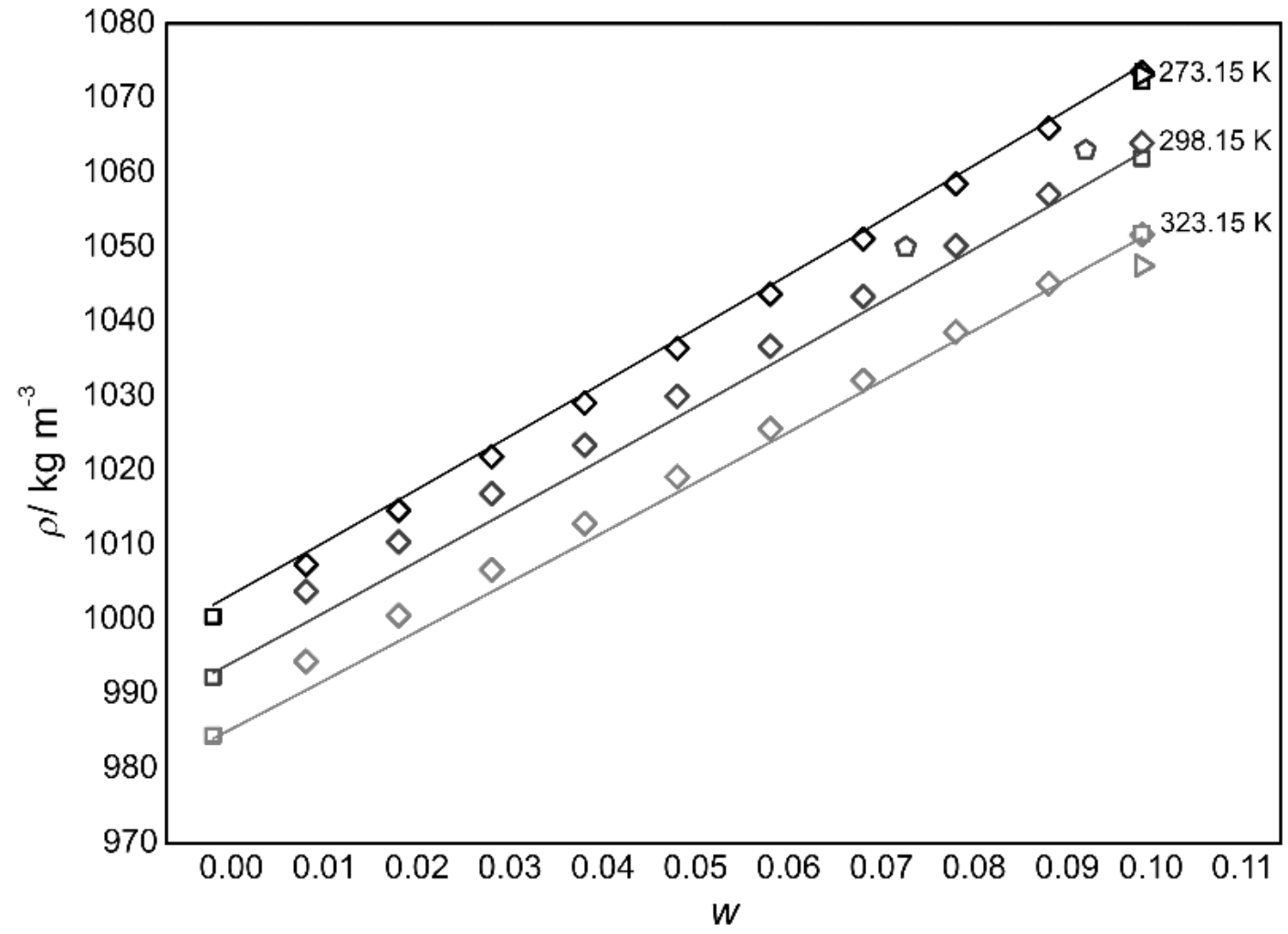

**Figure 4.** Extrapolation of the density parametrization: density values $\rho$ at low mass fraction range $w$ (0 to 0.1) in the temperature range $T$ (273.15 to 323.15 K). ◇ Washburn[44], △ Fasullo[46], ▷ Maksimova[52], ⬠ Liu[59] and - This work, Eq(1).

Moreover, we can affirm that the maximum percent deviation between our formula and the experimental values of literature is less than 1 % for concentrations lower than 0.53 $w$.[46, 52] If we take a greater percent deviation, like 1.5 %, the correspondence will go up to 0.55 $w$ concentration. In the case of Washburn[44] the correspondence is greater than for the other experimental values obtaining a maximum percent deviation of less than 1 % for concentrations lower than 0.57 $w$, and 1.5 % up to 0.63 $w$. Regarding mass fractions higher than 0.65 $w$, Figure 5 shows the nonlinear behaviour of the density, which maximum is around 0.96 to 0.97 *w*. Our linear fitting equation does not fit well in that region, all in all, is not the aim of the equation.

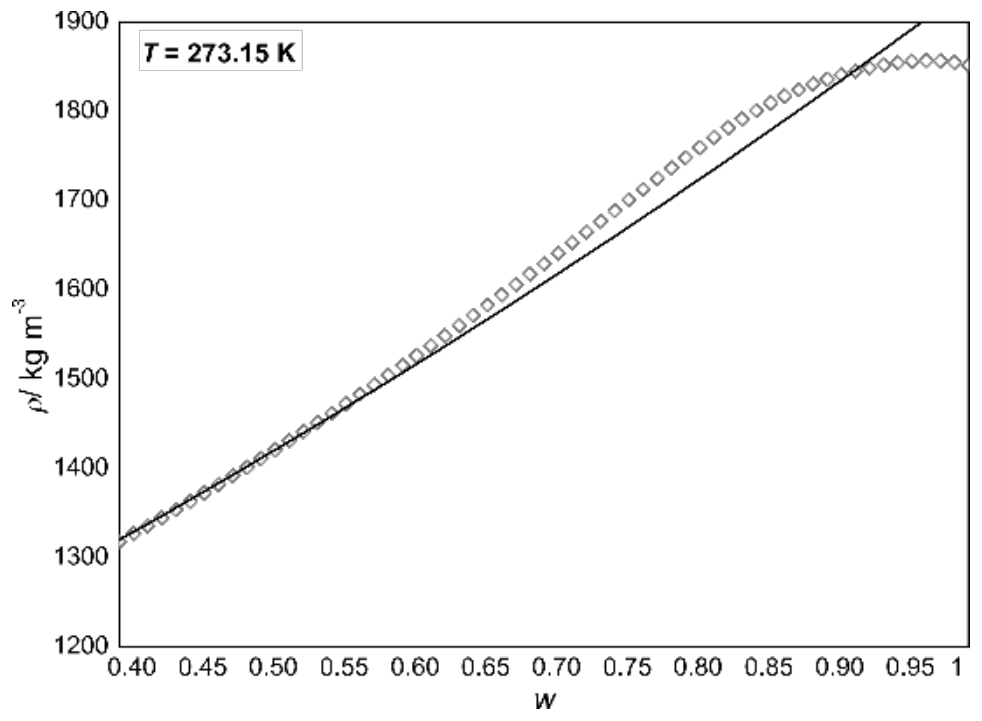

**Figure 5.** Extrapolation of the density parametrization: density values $\rho$ at mass fraction *w* (40 to100) at $T$ = 273.15 K. ◇ Washburn[44] and - This work, Eq. (1).

### 2.3.3. High temperature range, 333.15 < *T* < 373.15 K

As for high temperatures, density values obtained with the fitted equation fit well with the values of the bibliography[44, 46, 52, 59] in between 0 to 0.5 *w*. The greater deviation of the equation with respect to the experimental values appears at 0 *w* (pure water). As an example, Figure 6 shows the experimental density values at temperatures from 333.15 to 373.15 K for different concentrations and the fitted equation.

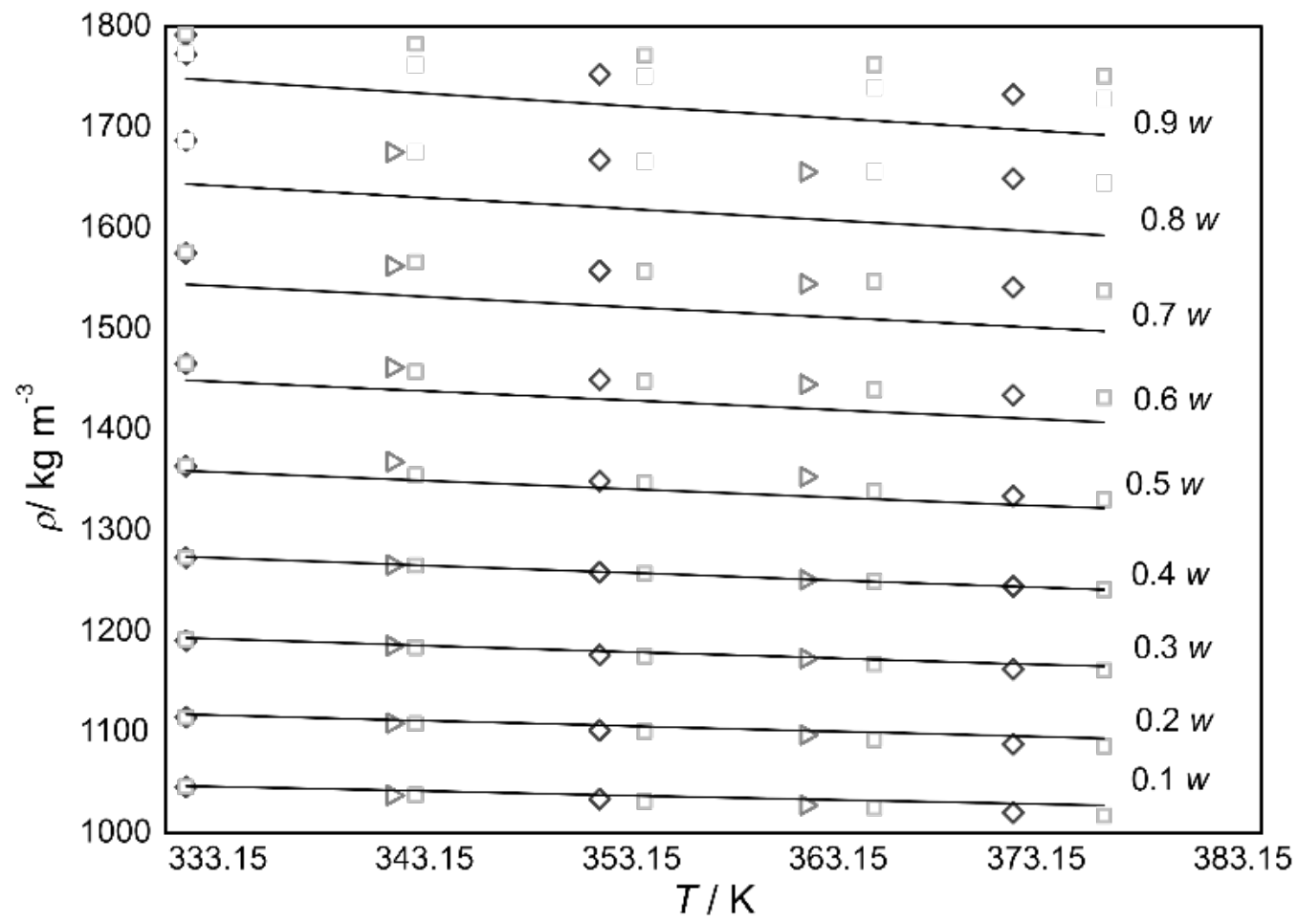

**Figure 6.** Extrapolation of the density parametrization: density values $\rho$ at high temperature range T (333.15 to 378.15 K) in the full mass fraction range *w* (0.1 to 0.9). ◇ Washburn[44], △ Fasullo[46], ▷ Maksimova[52] and - This work, Eq(1).

The maximum percent deviation increases with higher temperatures of electrolyte. Anyway, for the measurements[44, 46, 52, 59] the maximum percent deviation is lower than 1 % in 0 to 0.5 *w* for the 333.15 to 373.15 K range. In higher mass fractions than 0.6 *w*, the standard deviation is inferior than 5 %[44].

### 2.3.4. Conclusions of extrapolation of the fitting

We can confirm that the polynomial formula fitted in this work (equation (1) and Table 4) shows a maximum percent deviation with experimental values of less than 1 % in the next regions: {*w* (0 to 0.5); *T* (273.15 to 373.15 K)} and {*w* (0.12 to 0.67); *T* (221.15 to 273.15 K)}. Therefore, the equation is valid in the mentioned working ranges.

## 3. Thermal and mass expansion coefficients

In this section, thermal and mass expansion coefficients of sulfuric acid-water electrolyte have been analysed and compared with bibliography values.

### 3.1. Coefficient of thermal expansion

The coefficient of thermal expansion is the tendency of matter to change in volume in response to a change of temperature at atmospheric pressure. This coefficient at a fixed concentration is determined by the next formula:[74]

$$\alpha_T = -\frac{1}{\rho(T)}\left(\frac{\partial \rho}{\partial T}\right) \quad (2),$$

where α is the coefficient of thermal expansion.

The systematic analysis of $\alpha$ values can be used as important information to understand the influence of temperature on the intermolecular interactions in liquids. Indeed, the coefficient of thermal expansion is an important thermal property of a fluid, and is related to many different thermophysical quantities.[75] Temperature variations change the density of fluids, and $\alpha$ relates the two magnitudes. For example, this parameter is important in buoyancy, convection or lead-acid batteries, in which potential has measurable changes with temperature.[49]

The experimental evaluation of this coefficient for a concrete temperature range is normally performed using this formula.[49]

$$\rho_T = \rho(T_0)[1-\alpha(T-T_0)] \quad (3)$$

We have evaluated experimentally the coefficient of thermal expansion of the mixture in the region {*w* (0.1 to 0.4); *T* (273.15 to 333.15 K)} at atmospheric pressure. We have performed a linear fit of our experimental density values. The resulting values of the coefficient of thermal expansion of sulfuric acid-water electrolyte appear in Table 6.

**Table 6.** Experimental values of the coefficient of thermal expansion of the sulfuric acid-water solutions ($\alpha \cdot (10^{-4}\ K^{-1})$). In each point, for a specific mass fraction *w*, we have taken α ±10 K range of temperatures T, except for the 298.15 K value, in which the temperature range has been ±15 K.

| *w* | 283.15 K | 293.15 K | 298.15 K | 303.15 K | 313.15 K | 323.15 K |
|---|---|---|---|---|---|---|
| 0.097 | 3.2624 | 3.7691 | 3.9326 | 4.1228 | 4.1790 | 4.7997 |
| 0.115 | 3.7003 | 4.0970 | 4.1050 | 4.1646 | 4.5201 | 4.7858 |
| 0.137 | 4.0784 | 4.4005 | 4.5559 | 4.7203 | 4.3442 | 4.6173 |
| 0.154 | 4.3623 | 4.6066 | 4.7309 | 4.8573 | 4.3072 | 4.2344 |
| 0.173 | 4.6648 | 4.8581 | 4.9303 | 5.0059 | 4.9750 | 5.1986 |
| 0.192 | 4.9609 | 5.0477 | 4.8901 | 4.7771 | 4.5788 | 4.7198 |
| 0.211 | 5.1551 | 5.2380 | 5.2865 | 5.3321 | 5.3371 | 4.7518 |
| 0.230 | 5.3744 | 5.3727 | 5.4134 | 5.4368 | 4.2911 | 5.1014 |
| 0.241 | 5.4595 | 5.4875 | 5.3200 | 5.2006 | 5.2665 | 5.6618 |
| 0.269 | 6.0705 | 5.6060 | 5.6183 | 5.6150 | 5.6715 | 5.3498 |
| 0.288 | 5.7200 | 5.7049 | 5.6739 | 5.6441 | 5.6727 | 5.3431 |
| 0.307 | 5.6365 | 5.4367 | 5.4092 | 5.4808 | 5.4640 | 5.4216 |
| 0.327 | 5.8055 | 5.7170 | 5.7647 | 5.7751 | 5.8105 | 5.3697 |
| 0.346 | 5.7989 | 5.7744 | 5.7035 | 5.6457 | 5.6419 | 5.0515 |
| 0.365 | 5.7879 | 5.7552 | 5.7388 | 5.7185 | 5.7089 | 5.7072 |
| 0.384 | 5.7588 | 5.7409 | 5.6981 | 5.6611 | 5.6331 | 5.1060 |
| 0.400 | 5.7478 | 5.7031 | 5.6845 | 5.6642 | 5.5360 | 5.5252 |

In the analysed range, the minimum value is 3.2624 $10^{-4}$ $K^{-1}$ (283.15 K and 0.0966 *w*), and the maximum 6.0705 $10^{-4}$ $K^{-1}$ (283.15 K and 0.2688 *w*). The values of the coefficient increase with concentration. At low concentrations, the coefficient increases with the temperature, but at higher concentrations, it remains nearly constant.

To our knowledge, there are a few experimental values of the coefficient of thermal expansion of this electrolyte in the literature. For example, Bode[49] uses the densities of Washburn[44] and the linear relation of equation (3) to calculate the coefficient of thermal expansion between 273.15 K and 323.15 K in the entire concentration range 0 to 1 *w*. Bode gives a value for $\alpha$ of 6 $10^{-4}$ $K^{-1}$ for a charged lead-acid battery at 298.15 K (1250−1280 kg $m^{-3}$). For similar conditions, 298.15 K and 1269.6 kg $m^{-3}$, our value is 5.7388 $10^{-4}$ $K^{-1}$ (see Table 4 and Table 6). Salkind *et al.* calculate more values of the coefficient of thermal expansion of the electrolyte between 288.15 and 298.15 K.[57] They theoretically make a relation between density and the coefficient of thermal expansion by means of equation (3). At 298.15 K and 1146 kg $m^{-3}$, Salkind *et al*. get a value for $\alpha$ of 6 $10^{-4}$ $K^{-1}$, and our value for similar conditions is 5.2865 $10^{-4}$ $K^{-1}$ (298.15 K and 1144.5 kg $m^{-3}$). Different values for $\alpha$ can be found in the literature,[54-55, 76-78] and our results are in good agreement with all the analysed works.

### 3.2. Coefficient of mass expansion

Another important thermophysical property of the sulfuric acid-water electrolyte is the coefficient of mass expansion, $\beta$. For example, in the normal operation of the lead-acid batteries, the electrolyte concentration changes due to the electrochemical reactions related to charge and discharge processes. As a result, important density changes appear close to the electrodes.

The coefficient of mass expansion $\beta$, of the electrolyte at atmospheric pressure and temperature can be defined as follows:[74]

$$\beta_w = \left[ \frac{1}{\rho(w)} \left( \frac{\partial \rho}{\partial w} \right) \right] \qquad (4).$$

By means of the coefficient of mass expansion, we can calculate the density in the neighbourhood of the measured points at constant temperature. We assume that the density of the electrolyte has a linear relation with the concentration at constant temperature. Therefore, we can calculate the mass expansion coefficient at constant temperature using this linear relation:

$$\rho_w = \rho(w_0)[1+\beta(w-w_0)] \quad (5).$$

We have determined the coefficient of mass expansion using the measurements of the density of Table 4. For the calculation of each point, we have taken a ± 0.02 *w* range around the point (three values of the density with the same temperature). Then, we have fitted a linear function of the concentration with them. Afterwards, we have applied equation (5) to this linear fit of the density. Table 7 shows the results of the calculations.

**Table 7.** Experimental values of the coefficient of mass expansion of the sulfuric acid-water solutions $\beta$ ($10^{-3}$), at different temperatures T and mass fractions *w*.

| *w* | 273.15 K | 283.15 K | 293.15 K | 298.15 K | 303.15 K | 313.15 K | 323.15 K | 333.15 K |
|---|---|---|---|---|---|---|---|---|
| 0.115 | 7.1282 | 6.9219 | 6.7543 | 6.6942 | 6.6342 | 6.4927 | 6.4877 | 6.4607 |
| 0.137 | 7.0248 | 6.8246 | 6.6614 | 6.6031 | 6.5450 | 6.4105 | 6.3981 | 6.3768 |
| 0.154 | 6.9556 | 6.7778 | 6.6273 | 6.5702 | 6.5314 | 6.4725 | 6.4502 | 6.4169 |
| 0.173 | 6.9357 | 6.7715 | 6.6430 | 6.5876 | 6.5501 | 6.5111 | 6.3071 | 6.4423 |
| 0.192 | 7.1035 | 6.9630 | 6.8483 | 6.8040 | 6.7678 | 6.6828 | 6.7317 | 6.5486 |
| 0.211 | 7.1329 | 7.0047 | 6.9105 | 6.8675 | 6.8244 | 6.7820 | 6.8180 | 6.6422 |
| 0.230 | 7.1017 | 6.9006 | 6.8264 | 6.8013 | 6.7594 | 6.6491 | 6.5215 | 6.4368 |
| 0.241 | 7.0937 | 6.9712 | 6.8987 | 6.8749 | 6.8509 | 6.8072 | 6.6375 | 6.6843 |
| 0.269 | 6.8544 | 6.7658 | 6.7791 | 6.7474 | 6.7328 | 6.7202 | 6.4776 | 6.6983 |
| 0.288 | 6.7886 | 6.7618 | 6.7510 | 6.7453 | 6.7313 | 6.6605 | 6.6484 | 6.7067 |
| 0.307 | 6.6959 | 6.6722 | 6.6577 | 6.6524 | 6.6388 | 6.5664 | 6.5543 | 6.6130 |
| 0.327 | 6.6122 | 6.5866 | 6.5767 | 6.5706 | 6.5572 | 6.4901 | 6.4782 | 6.5354 |
| 0.346 | 6.5321 | 6.5069 | 6.4970 | 6.4918 | 6.4785 | 6.4100 | 6.3983 | 6.4506 |
| 0.365 | 6.4517 | 6.4267 | 6.4168 | 6.4116 | 6.3984 | 6.3318 | 6.3200 | 6.3804 |
| 0.384 | 6.3645 | 6.3396 | 6.3297 | 6.3245 | 6.3115 | 6.2451 | 6.2333 | 6.2854 |

As we have shown in section 2, electrolyte density is related to concentration (see Figure 1). Therefore, the coefficient of mass expansion maintains approximately a constant value, oscillating from a maximum value of 7.1329 $10^{-3}$ to a minimum value of 6.2333 $10^{-3}$. In this range, its average value is 6.6382 $10^{-3}$ (see Table 7).

To our knowledge, there are no experimental values of the coefficient of mass expansion of sulfuric acid-water mixtures. Consequently, it is not possible to check our results with other researches.

### 3.3. Density fitting based on thermal and mass expansion coefficients

In order to conclude the analysis of the thermal and mass expansion coefficients, considering a linear variation of the density with the temperature and concentration, thus, the linear Boussinesq approximation equation has been used.

$$\rho = \rho_0[1-\alpha(T-T_0)+\beta(w-w_0)] \qquad (6),$$

where $T_0$ corresponds to a reference temperature of 298.15 K, $w_0$ is the reference mass fraction of 0.24 and $\rho_0$ the density in those temperature and mass fraction.

A correlation between the density experimental values of this work and the equation 6 has been plotted in Figure 7. Thermal and mass expansion coefficients of Table 6 and Table 7 has been used. The validity of equation 6 is limited in high mass fractions, all in all, the maximum error is less than 1 %.

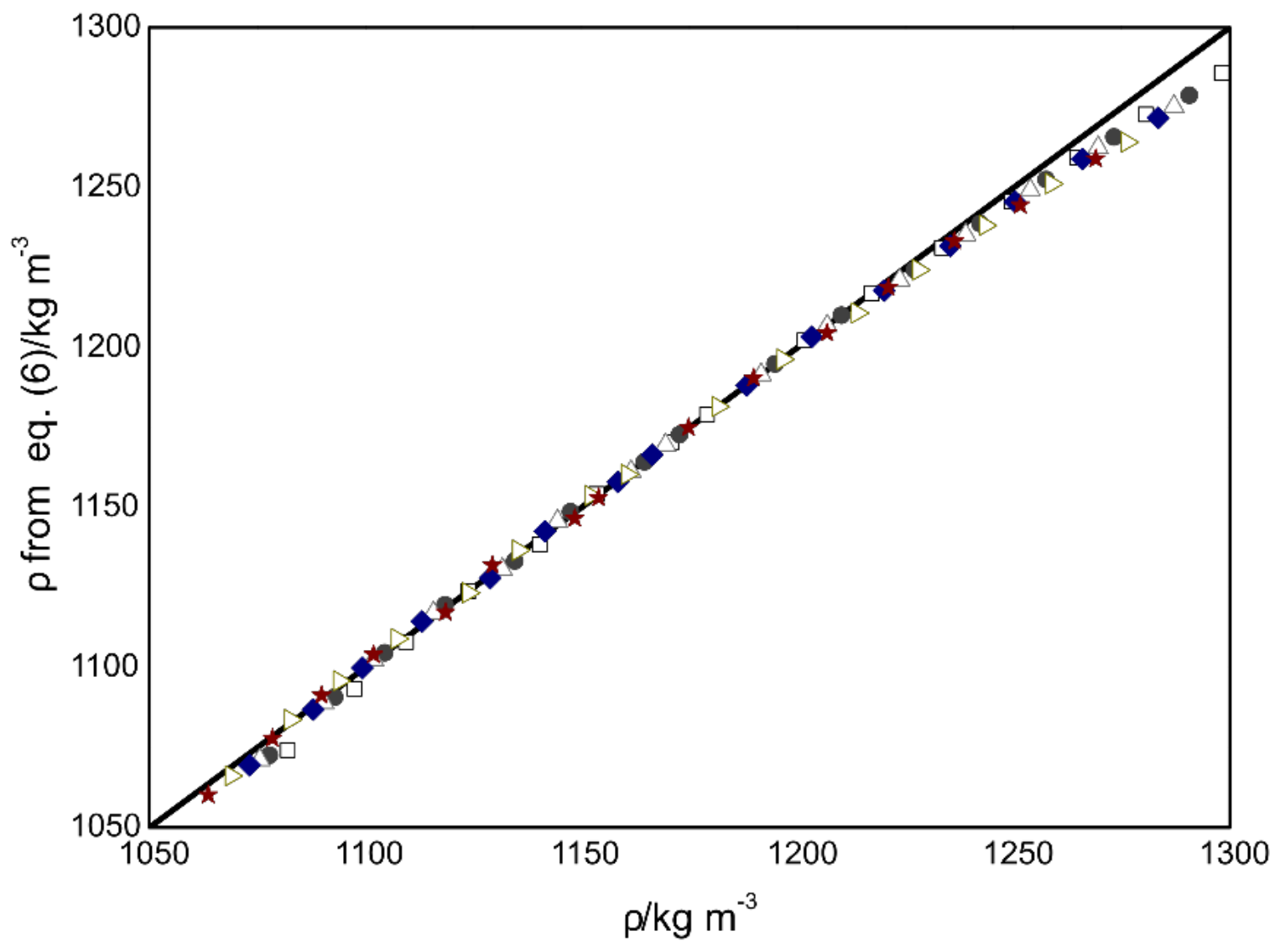

**Figure 7.** Correlation between experimental density and density fitting based on thermal and mass expansion coefficients from eq. (6). □ 283.15 K, ● 293.15 K, △ 298.15 K, ◆ 303.15 K, ▷ 313.15 K and ★ 323.15 K.

## 4. Viscosity

Investigations on the viscosity of sulfuric acid mixtures have been performed since the early 1900s. Some of the existing data of viscosity of those mixtures are presented in Table 8, giving extra information of the employed measuring techniques.

**Table 8.** Dynamic viscosity of the sulfuric acid-water solutions in the literature chronologically ordered.

| First author | Year | T / K [number of values or $\Delta T$ from previous temperature in K] | $w$ [number of values or $\Delta w$ in %] | μ techniques |
|---|---|---|---|---|
| Rhodes[43] | 1923 | 273.15 to 348.15 [25 K] | 0 to 0.996 [20] | Viscometer (cylindrical glass bulb with 2 $10^{-4}$ $m^3$ capacity, glass capillary tube of 0.08 m and 0.001 m internal diameter). |
| Vinal[79] | 1933 | 223.15 to 303.15 | 0.1 to 0.5 [10 $w$] | Timing the discharge of a fixed volume solution on a pyrex glass of ratio 100. |
| Das[80] | 1997 | 218.15 to 298.15 | >0.66 | Cannon-Ubbelohde suspended level viscometer |
| G. G. Aseyev[81] | 1998 | 233.15 to 358.15 [5 K] *1 | 0.02 to 0.76 [2 $w$] | Not measured. Multiple sources. |
| Walrafen[69] | 2000 | - | - | Not measured. Rhodes[46] |

*1 Note that Aseyev has not measured all the mass fractions in all the temperatures. At low temperatures, the measured mass fractions are: {*w* (0.1 to 0.22, T (> 263.15 K); *w* (0.24), T (>258.15 K); *w* (0.26 and 0.6 to 0.66), T (> 253.15 K); *w* (0.28 and 0.68), T (> 248.15 K) and *w* (0.70), T (> 243.15 K)}.

Our viscosity measurements were performed with Haake Falling Ball Viscometer, which has a precision error of ± 1 %. The temperature was maintained at 298.15 ± 0.1 K. In order to ensure the reproducibility of the measurements, each sample was measured eight times. The experimental results of the dynamic viscosity are shown in Table 9.

**Table 9.** Measured dynamic viscosity μ of sulfuric acid-water electrolyte samples at 298.15 K for different mass fractions *w*.

| *w* | μ / Pa s | *w* | μ / Pa s |
|---|---|---|---|
| 0.097 | 1072.76 | 0.230 | 1455.33 |
| 0.115 | 1110.00 | 0.241 | 1481.07 |
| 0.137 | 1160.58 | 0.269 | 1593.59 |
| 0.154 | 1221.21 | 0.288 | 1691.71 |
| 0.173 | 1238.55 | 0.307 | 1778.38 |
| 0.192 | 1322.39 | 0.327 | 1875.41 |
| 0.211 | 1360.69 | 0.346 | 2060.22 |

In the Figure 8, Rhodes[43], Vinal[79] and Georgievich[81] viscosity values are graphically compared with our measurements (presented in Table 9), presenting the same tendency in the whole concentration range for the studied temperature (298.15 K).

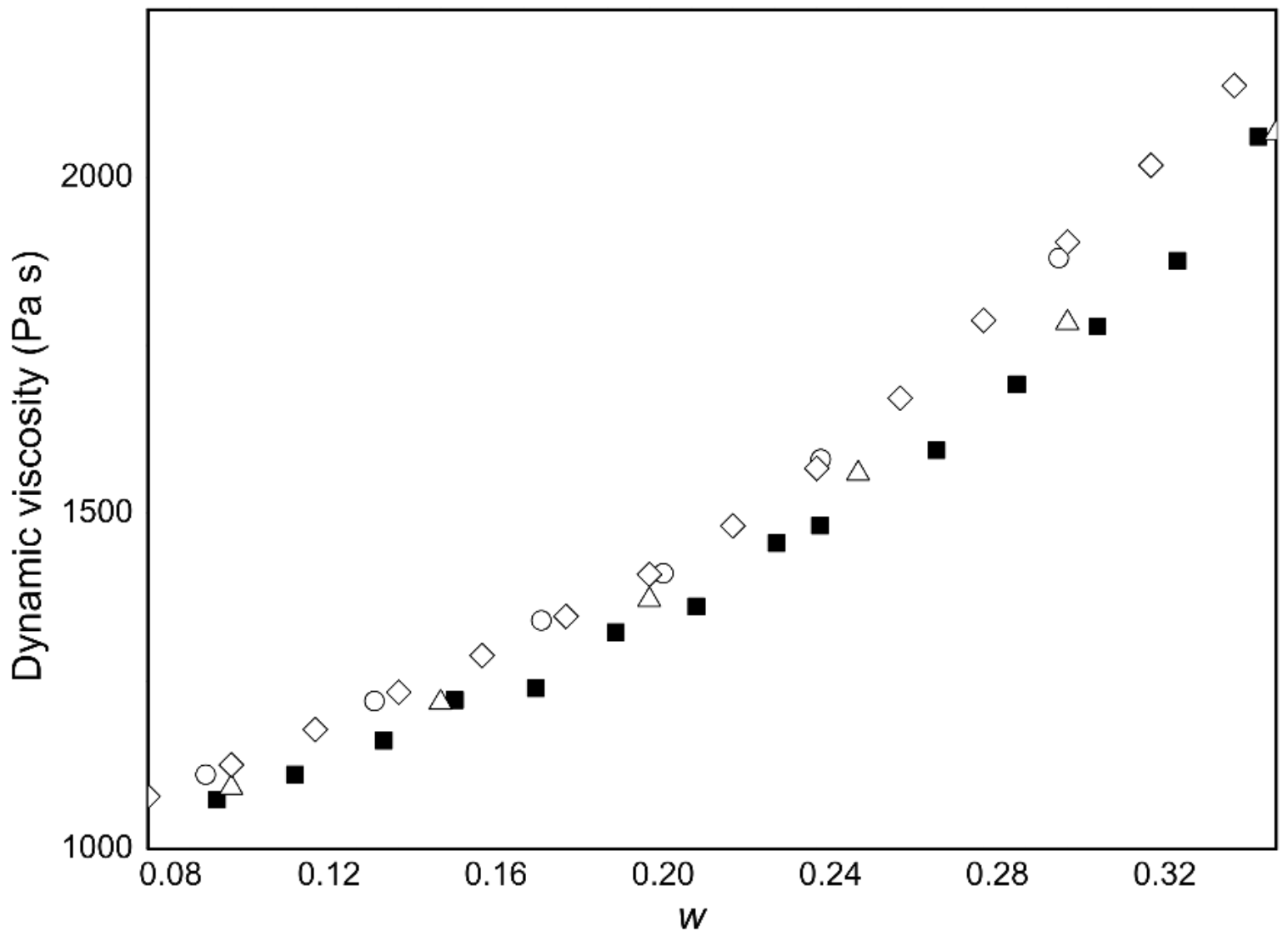

**Figure 8.** Comparison of experimental viscosity values of this work μ and bibliography at 298.15 K. ○ Rhodes[46], △ Vinal[79], ◇ G. G. Aseyev[81], ■ This work.

A second order polynomial equation is used for fitting our results, with a root mean square error of 0.996.

$$\mu = 0.0009w^2 - 0.0045w + 1.0421 \quad (7).$$

In eq. (7), a fitting equation for the dynamic viscosity μ for different mass fractions *w* is presented. Moreover, experimental values, compared to literature values, in all the cases, deviates a maximum of 10 %, appreciated at high mass fractions.

## 5. Index of refraction

The concentration measurement of little amounts of electrolyte is very important in different industrial applications and nature phenomena (e.g. lead acid battery monitoring[82] and atmospheric nucleation processes[83]). Indeed, the measurement of the refractive index can be used as an indirect method to determine the concentration of the electrolyte. For this reason, the characterization of the refraction index of sulfuric acid-water electrolyte for different concentrations is interesting.

Although the properties of sulfuric acid-water electrolyte have been profusely studied from long time ago, the refractive index of this mixture has not been studied so deeply. Initial measurements of refractive index of sulfuric acid-water electrolyte started in 1974.[84] Table 10 shows a review of refractive index data of the literature, with the more remarkable information about measurements.

**Table 10.** Refraction index of the sulfuric acid-water solutions in the literature chronologically ordered.

| First author | Year | T / K | *w* | $\lambda$ /nm |
|---|---|---|---|---|
| Querry[84] | 1974 | 300.15 | 0.25 | 2,000 to 20,000 |
| Remsberg[85] | 1974 | 296.15 | 0.75, 0.9 | 6,000 to 13,000 |
| Palmer[86] | 1975 | 300.15 | 0.25, 0.38, 0.5, 0.75, 0.845, 0.956 | 357 to 2,500 |
| Beyer[87] | 1996 | 298.15 | 0, 0.398, 0.523, 0.63, 0.787, 0.886 | 214, 254, 308, 313, 365 |
| Born[88] | 1999 | 288.15 | 0, 0.1998, 0.3976, 0.5998, 0.801, 1 | 589 |
| Niedziela[89] | 1999 | 200.15 to 300.15 | 0.32 to 0.87 | 2,632 |
| Biermann[90] | 2000 | 183.15 to 293.15 | ≤ 0.8 | 357 to 2,500 |
| Krieger[91] | 2000 | 213.15 to 303.15 | 0, 0.1004, 0.1741, 0.217, 0.351, 0.4, 0.434, 0.5015, 0.651 | 351.0, 533.5, 632.9, 782.6 |
| Lund-Myhre[92] | 2003 | 220.15 to 300.15 | 0.12 to 0.81 | ≥ 1,333 |

A RFM340 series refractometer of Bellingham and Stanley has been used to perform the refraction index measurements (wavelength of 589.3 nm). We have used electrolyte samples of 1 ml. The temperature of the sample remained constant at 293.15 K during measurements. The experimental values of the refractive index are shown in Table 11.

**Table 11.** Measured refractive index of sulfuric acid-water electrolyte samples at 293.15 K ($\lambda$ = 589.3 nm).

| *w* | *n* | *w* | *n* |
|---|---|---|---|
| 0.000 | 1.33297 | 0.260 | 1.36566 |
| 0.099 | 1.34523 | 0.280 | 1.36812 |
| 0.121 | 1.34761 | 0.300 | 1.37054 |
| 0.150 | 1.35162 | 0.321 | 1.37277 |
| 0.161 | 1.35276 | 0.339 | 1.37596 |
| 0.179 | 1.35510 | 0.362 | 1.37853 |
| 0.200 | 1.35744 | 0.381 | 1.38042 |
| 0.221 | 1.36047 | 0.400 | 1.38311 |
| 0.240 | 1.36249 | 0.423 | 1.38683 |

Our measured values of the refractive index show a linear behaviour with mass fraction. Figure 9 a) shows this linear behaviour of the refractive index Therefore, we can correlate linearly refractive index with mass fraction, as shown in Eqs (8):

$$n(w) = 1.3 \cdot 10^{-3} w + 1.3325 \quad (8)$$

Combining Eq (1) and Eq (8) it is possible to find a similar relation of the refractive index and the density, which has a similar linear behaviour. These linear relations are valid only for concentrations lower than 0.4 *w*, as proved by Beyer[87]. For this mass fraction range, our measurements are in concordance with those values of literature (see Figure 9 b)).

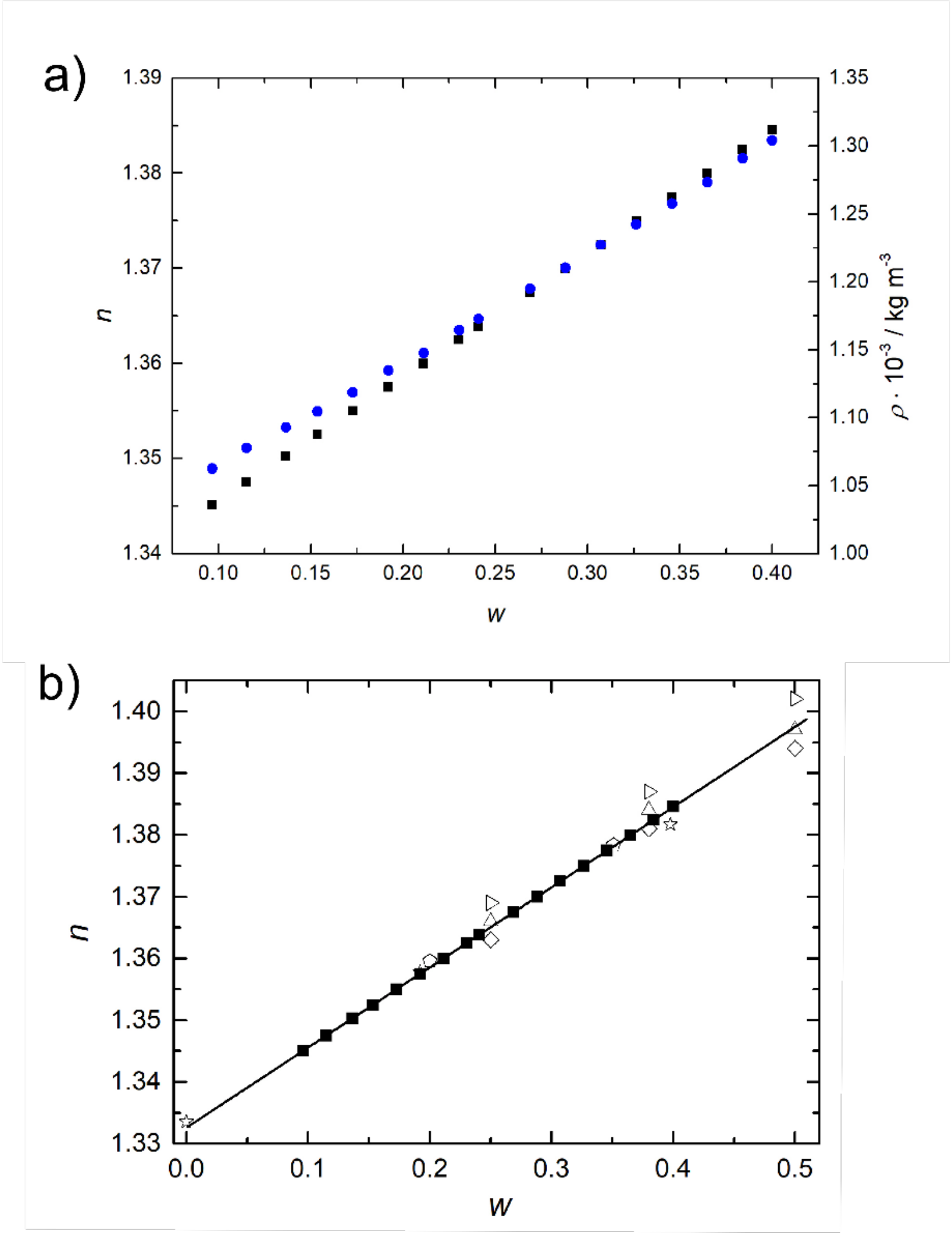

**Figure 9.** Electrolyte refractive index. a) Linear relations between our measured refractive index and the concentration and density of the electrolyte at 293.15 K. ● *n* and ■ *ρ*; b) Comparison between our measured refractive index values at 293.15 K and those of the literature. ◇ Palmer[86] 300.15 K / 702 nm, △ Palmer[86] 300.15 K / 556 nm, ▷ Palmer[86] 300.15 K / 449 nm, ☆ Born[88] 288 K / 589 nm, ⬠ Krieger[91] 294 K / 533.5 nm, ■ This work.

## 6. Conclusions

Sulfuric acid-water electrolyte is a chemical of great industrial importance. The aim of this work is to compare and contrast literature values of the thermophysical properties of this electrolyte with newly measured values, providing an historic overview of those properties. For this purpose, we have collected and compared literature values of density, coefficient of thermal and mass expansion, dynamic viscosity and refractive index of this electrolyte. Moreover, we have collected also density parametrizations from literature to compare with a new and more simple equation developed in this work.

The exhaustive review of density values shows that there is plenty of measured data available in the literature. The most used resource is that of the International Critical Tables. However, some measurements have increased the information for dilute solutions and low temperatures. We have chosen five of the literature data sets to compare with our density measurements and a new optimized parametrization of the density developed from our measurements. The extrapolation of the new parametrization has defined the working range of the equation in these regions: {$w$ (0 to 0.5); T (273.15 to 373.15 K)} and {$w$ (0.12 to 0.67); T (221.15 to 273.15 K)}. We have found a maximum difference between the parametrization and the experimental densities of less than 1 %.

We have also calculated the coefficients of thermal and mass expansion of the sulfuric acid-water mixtures using our measurements. There are few values of the thermal expansion coefficient in the literature, but they show a good agreement with our values. There are not reported values of the mass expansion coefficient. However, in order to check our values, we have correlated them using the Boussinesq approximation and the agreement is good.

We have also measured two other important properties of the electrolyte: the dynamic viscosity (298.15 K) and the refraction index (293.15 K). In both cases, the new measured values agree with literature ones.

**Acknowledgements**

The authors acknowledge D. Soler and E. Zarate for help in the development of the density parametrization. Research Group Program (IT 1009-16) and μ4F (KK-2017/00089) of the Basque Government.